\documentclass{article}
\usepackage{spconf,amsmath,graphicx}
\usepackage[table,xcdraw]{xcolor}
\usepackage[ruled,vlined]{algorithm2e}
\usepackage{multirow}
\graphicspath{{images/}}

\newcommand{\firsttable}
{% no arguments
\begin{table}[]
\resizebox{\columnwidth}{!}{%
\begin{tabular}{|c|l|c|c|c|c|c|}
\hline
\rowcolor[HTML]{FFFFFF} 
 &
  \multicolumn{1}{c|}{\cellcolor[HTML]{FFFFFF}Material} &
  \begin{tabular}[c]{@{}c@{}}Density\\ ($gr/cm^3$)\end{tabular} &
  \begin{tabular}[c]{@{}c@{}}\# of views \\ {[}Input, Ref.{]}\end{tabular} &
  \begin{tabular}[c]{@{}c@{}}Short-Scan\\ {[}Input, Ref.{]}\end{tabular} &
  \begin{tabular}[c]{@{}c@{}}Integration Time (s)\\ $\times$ No. Image Averaging\end{tabular} &
  \begin{tabular}[c]{@{}c@{}}{[}Beam Hardening Filter, \\ Current (mA), Voltage (kV) {]}\end{tabular} \\ \hline
\rowcolor[HTML]{FFFC9E} 
\cellcolor[HTML]{FFFC9E} &
  AlCe &
  3 &
  {[}145, 580{]} &
  {[}True, True{]} &
  1 $\times$ 4 &
  {[}0.25mm Cu, 130,225{]} \\ \cline{2-7} 
\rowcolor[HTML]{FFFC9E} 
\cellcolor[HTML]{FFFC9E} &
  Stainless steel &
  7.9 &
  {[}334, 1000{]} &
  {[}False, False{]} &
  1 $\times$ 8 &
  {[}0.25mm Cu, 130,225{]} \\ \cline{2-7} 
\rowcolor[HTML]{FFFC9E} 
\multirow{-3}{*}{\cellcolor[HTML]{FFFC9E}\begin{tabular}[c]{@{}c@{}}Training \\ and \\ Testing (In-Dist.)\end{tabular}} &
  Inconel 718 &
  8.2 &
  {[}166, 500{]} &
  {[}True, True{]} &
  1 $\times$ 4 &
  {[}0.25mm Cu, 130,225{]} \\ \hline
\rowcolor[HTML]{68CBD0} 
\cellcolor[HTML]{68CBD0} &
  \multicolumn{1}{c|}{\cellcolor[HTML]{68CBD0}NiCo} &
  8.8 &
  [334, 1000] &
  [False, False] &
  [1 $\times$ 8] &
  {[}0.5mm Cu, 130,225{]} \\ \cline{2-7} 
\rowcolor[HTML]{68CBD0} 
\multirow{-2}{*}{\cellcolor[HTML]{68CBD0}\begin{tabular}[c]{@{}c@{}}Testing \\ (OOD)\end{tabular}} &
  \multicolumn{1}{c|}{\cellcolor[HTML]{68CBD0}Inconel 718} &
  8.2 &
  [145, -] &
  [True, -] &
  [1 $\times$ 8] &
  {[}0.25mm Cu, 130,225{]} \\ \hline
\end{tabular}%
}
\caption{Description of measurement setting and materials used in this work. 
Please note that the test Inconel 718 is OOD, because it was printed with a different printing system than the the one used during training, and scanned with a different setting.
}
\vspace{-0.25cm}
\label{Tab:Meas_setting}
\end{table}
}

\newcommand{\secondtable}
{% no arguments
\begin{table}[htb!]
\begin{center}
\footnotesize %\small
\begin{tabular}{|c | c | c | c | c|} 
 \hline
Image & Scan type & BHC & Sparse-view correc. & Recon.\\
 \hline\hline
 Uncorrected & Sparse & - & - & FDK \\ 
 \hline
 Reference & Dense & CAD-phys & - & MBIR \\
 \hline
 Our Method & Sparse & BHCN & DLMBIR & FDK \\
 \hline
\end{tabular}
\caption{Description of images being compared in the results section.}
\label{Tab:Image_description}
\end{center}
\vspace{-0.5cm}
\end{table}
}

\newcommand{\thirdtable}
{\begin{table}[htb!]
\begin{center}
\begin{tabular}{| c | c | c | c | c |} 
 \hline
& Image & Steel & AlCe & NiCo\\
 \hline\hline
 \multirow{2}{*}{PSNR} & Uncorrected & 32.24 & 37.72 & 34.12\\ 
 \cline{2-5}
 & Our Method & 42.23 & 43.02 & 40.97\\ 
 \hline\hline
 \multirow{2}{*}{SSIM} & Uncorrected & .9845 & .9816 & .9804\\ 
 \cline{2-5}
 & Our Method & .9914 & .9929 & .9850\\ 
 \hline
\end{tabular}
\caption{PSNR (dB) and SSIM w.r.t reference, averaged across the image slices for different alloys. 
The proposed method scores higher for all alloys in and out of distribution of the training data.
}
\vspace{-0.5cm}
\label{Tab:PSNR_SSIM}
\end{center}
\end{table}
}% file containing commands\input{Supplementary.tex}
\title{Deep learning based workflow for accelerated industrial X-ray Computed Tomography}
\name{ \parbox{\linewidth}{\centering{Obaidullah Rahman$^\dagger$, Singanallur V. Venkatakrishnan$^\dagger$, Luke Scime$^\dagger$, Paul Brackman$^{\star}$, Curtis Frederick$^{\star}$, Ryan Dehoff$^\dagger$, Vincent Paquit$^\dagger$, Amirkoushyar Ziabari$^\dagger$}}
\thanks{Corresponding author's email address: \textit{rahmano@ornl.gov}. 
This manuscript has been authored by UT-Battelle, LLC, under contract DE-AC05-00OR22725 with the US Department of Energy (DOE).Research sponsored by the US Department of Energy, Office of Energy Efficiency and Renewable Energy, Advanced Manufacturing Office and Technology Commercialization Fund (TCF-21-24881), under contract DE-AC05-00OR22725 with UT-Battelle, LLC.
The US government retains and the publisher, by accepting the article for publication, acknowledges that the US government retains a nonexclusive, paid-up, irrevocable, worldwide license to publish or reproduce the published form of this manuscript, or allow others to do so, for US government purposes. DOE will provide public access to these results of federally sponsored research in accordance with the DOE Public Access Plan (http://energy.gov/downloads/doe-public-access-plan).}}
\address{{$^{\dagger}$}Oak Ridge National Lab (ORNL), Oak Ridge, TN 37830\\ 
{$^{\star}$}Carl Zeiss Industrial Metrology, LLC, Maple Grove, MN 55369, USA}
\begin{document}
%\ninept
%%%%%%%%
\maketitle

\begin{abstract}
X-ray computed tomography (XCT) is an important tool for high-resolution non-destructive  characterization of additively-manufactured metal components. 
XCT reconstructions of metal components may have beam hardening artifacts such as cupping and streaking which makes reliable detection of flaws and defects challenging. 
Furthermore, traditional workflows based on using analytic reconstruction algorithms require a large number of projections for accurate characterization - leading to longer measurement times and hindering the adoption of XCT for in-line inspections.
In this paper, we introduce a new workflow based on the use of two neural networks to obtain high-quality accelerated reconstructions from sparse-view XCT scans of single material metal parts.
The first network, implemented using fully-connected layers, helps reduce the impact of BH in the projection data  without the need of any calibration or knowledge of the component material.
The second network, a convolutional neural network, maps a low-quality analytic 3D reconstruction to a high-quality reconstruction.
Using experimental data, we demonstrate that our method robustly generalizes across several alloys, and for a range of sparsity levels without any need for retraining the networks thereby enabling accurate and fast industrial XCT inspections.  
\end{abstract}

\vspace{-0.5cm}
\section{Introduction}
\label{sec:intro}
%what is the problem ? 
Additive manufacturing (AM), also known as 3D printing, is an important process for printing complex components that cannot be manufactured via traditional machining methods. 
In it known that during the AM process,  flaws such as pores (holes), cracks, and other defects,  may form in the printed objects~\cite{brennan2021defects, svetlizky2021directed}, that could compromise their performance, and therefore advanced characterization techniques are critical for qualification and certification the components.
XCT is a powerful method for non-destructive characterization of the flaws and defects in 3D at high-resolution. 
However, it is challenging to use XCT for the characterization of a large number of metal AM components.

%Why is it challenging ? 
Traditional workflows for XCT imaging based on using analytic reconstruction algorithms such as FDK~\cite{feldkamp1984practical} require a large  number  of  projections  --  leading  to  longer measurement times and incurring significant labor and cost.   
Attempts to accelerate the XCT scans by acquiring only a fraction of the typically-made measurements lead to strong artifacts in the reconstructions and produce incorrect characterization results. 
Model-based iterative reconstruction (MBIR) algorithms can help reduce scan time by producing high-quality reconstructions from sparse scans~\cite{ziabari2023enabling}; but they require significant computation making them infeasible to use  when characterizing hundreds of components. 
Another challenge with using industrial XCT systems for metal AM parts is that there can be significant artifacts in the reconstructions due to beam hardening (BH) from the interaction of a poly-chromatic X-ray source with dense metals. 
Several deep learning approaches have been developed to address these challenges ~\cite{ziabari2020beam,ziabari2022simurgh,ziabari2022high,ziabari2023enabling} for metal AM.
Those methods handle the beam hardening, sparse scan artifact reduction and denoising using the same network.
This could complicate the training process specially when materials of different density levels are being used during training, and also impact the performance of the network. 
In addition, the beam hardening correction for training data generation is material and X-ray spectrum dependant.

%What are we proposing ? 
In this paper we present a two-stage deep learning-based workflow that disentangles beam hardening correction from artifact reduction while enabling high-quality reconstruction from sparse scans. 
Specifically, we use a neural network (NN) to estimate the beam hardening related parameters, correct the acquired data based on the outputs of the NN, perform a FDK reconstruction on the corrected data and use a convolutional neural network to map the FDK reconstruction to a high-quality reconstruction. 
The first network is trained purely based on a model (requires no measurements/calibration data), allowing for a material-agnostic beam hardening correction. 
The second network is trained using a collection of densely-sampled measurements from different materials and scan settings. 
Furthermore, by removing the bias term from the convolutional neural network~\cite{mohan2019robust}, we attempt to improve the generalization. 
We shed light on potential generalizability of our approach, using several sparse-view XCT data sets of metal parts. 
We demonstrate that our approach produces high-quality reconstructions across a variety of samples relevant to AM applications, including those that are different from the type of samples used to train the neural networks. 

\section{Method}
\label{sec:method}
\begin{figure}[htb]
   \centering
   \centerline{\includegraphics[width=\linewidth]{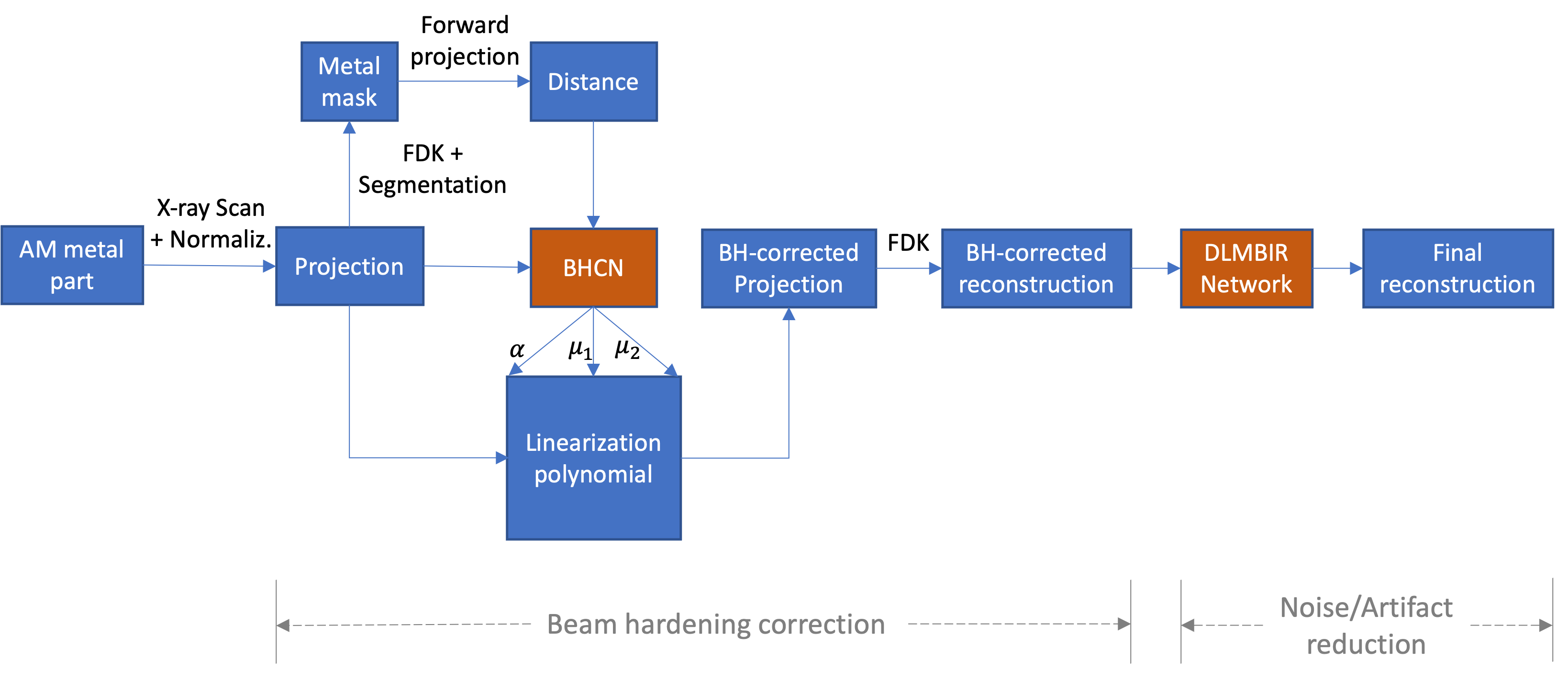}}
   \vspace{-0.5cm}
 \caption{Block diagram of our proposed workflow consisting of beam hardening correction based on a network (BHCN), analytic reconstruction (FDK) and a DNN for suppressing sparse-view artifacts and noise (DL-MBIR).
 }
 \vspace{-0.25cm}
 \label{fig:BHCN-DL-MBIR}
\end{figure}

In this section, we present details of the two-stage  neural network-based workflow for the BH artifacts and sparse-view artifact suppression (see Fig.~\ref{fig:BHCN-DL-MBIR}).
The first step is BH correction network (BHCN), which is a neural network trained on experimentally-driven synthesized values to correct for BH in the projection data.
The second step, is an artifact-reduction deep neural network to reduce noise and artifacts from the FDK reconstruction and to approximate MBIR-like dense-view reconstruction~\cite{ziabari20182, rahman2021mbir}.  
In the following subsections, we describe the two neural networks used in the proposed method.

\vspace{-0.2in}
\subsection{Beam hardening correction network (BHCN)}\label{sec:BHCN}
Our method, BHCN~\cite{rahman2023neural}, to correct for BH from single material scans consists of the following steps:
\vspace{-0.1in}
\begin{itemize}
\itemsep-0.4em 
\item{Use BHCN to map \textit{each} projection value ($p$) and estimated thickness value ($d$) to the BH model parameters}
\item{Average the model parameters estimated by the NN for all the ($p,d$) values}
\item{Use the averaged model parameters to compute linearization polynomial}
\end{itemize}
\vspace{-0.1in}
We train the proposed network solely on synthetically-generated data using bimodal energy model for BH developed by Van de Casteel et al.~\cite{van2002energy} who demonstrate that BH can be modeled using two dominant X-ray energies, $E_1$ and $E_2$. If $\mu_1$ and $\mu_2$ are the linear attenuation coefficients (LAC) of the material at these energies, the BH-affected projection is modeled as
\begin{equation}
    p_{bh} = \mu_2d + \ln{\frac{1+\alpha}{1+\alpha e^{-(\mu_1-\mu_2)d}}} \label{eq:BHeqn}
\end{equation}
where $d$ is the distance the X-ray beam traverses within the material (thickness). 
$\alpha$ represents the ratio of the product of approximate X-ray spectrum value and detector efficiency at $E_1 \ \text{and} \ E_2$. 
The ideal (BH-free) projection  varies linearly with distance and is given by
\begin{equation}
    p_{bhc} = \frac{\alpha\mu_1 + \mu_2}{1+\alpha}d \label{eq:BHCeqn}
\end{equation}
%
%\vspace{-.2cm}

\subsubsection{Training}
\begin{figure}[htb]
   \centering
   \centerline{\includegraphics[width=0.95\linewidth]{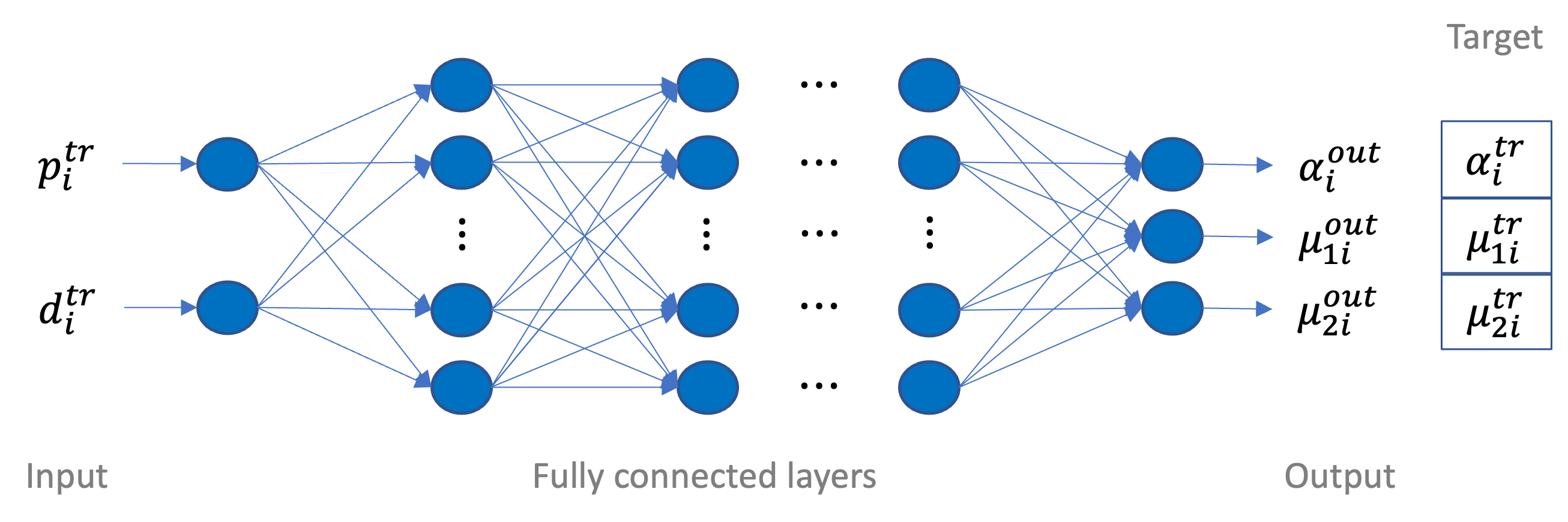}}
   \vspace{-0.5cm}
 \caption{Illustration of the BHCN architecture used for estimating the parameters of the beam hardening model in~\cite{van2002energy}. 
 The measured projection value and corresponding estimated thickness are provided as inputs and the network estimates the BH model parameters for each measurement in the CT scan. 
 The predicted model parameters are then averaged to obtain a single set of $\alpha, \mu_1$ and $\mu_2$ values which are then used to compute a linearization polynomial.
 } 
%  \vspace{-0.5cm}
 \label{fig:FCN}
\end{figure}

%\vspace{-.2cm}
We used a NN of fully connected layers, which we call beam hardening correction network (BHCN) shown in Fig.~\ref{fig:FCN}.
The input layer consists of 2 nodes for the BH affected projection value and associated thickness; and the output layer consists of 3 nodes for the parameters of the model in \eqref{eq:BHeqn}.
Samples of $d^{tr},\ \alpha^{tr}, \mu_1^{tr},\ \text{and }\mu_2^{tr}$ are uniformly drawn from a realistic range, and used to calculate $p^{tr}$ using Eq.~\ref{eq:BHeqn}. 
This range was inferred from past measurements of the densest and lightest alloys in our facility and going about $20\%$ beyond and $15\%$ below their range. 
Therefore we input the pair ($p^{tr}$, $d^{tr}$) to BHCN and train it by minimizing the error
between the output ($\alpha^{out}$, $\mu_1^{out}$, $\mu_2^{out}$) and the target ($\alpha^{tr}$, $\mu_1^{tr}$, $\mu_2^{tr}$).
%\vspace{-.2cm}

\subsubsection{Inference}
\label{sec:BHCN_Inference}
%\vspace{-.2cm}
To obtain the parameters of Van de Casteel model~\cite{van2002energy} from the BHCN, we first reconstruct the measured data using the FDK algorithm~\cite{feldkamp1984practical}.
Next, we perform a binary segmentation of this reconstruction using Otsu's algorithm~\cite{otsu1979threshold} and forward project it to obtain an estimate of the thickness corresponding to each measured projection.  
Then, the BH-affected projection and thickness estimates are fed into the BHCN to obtain estimates of vectors $\alpha$, $\mu_1$ and $\mu_2$.
The mean of the outputs corresponding to each input is used to compute the final linearization polynomial.
The projections are then corrected for BH followed by the use of the FDK to obtain a 3D reconstruction. 

\vspace{-0.25cm}
\subsection{Deep Learning MBIR (DL-MBIR)}
\begin{figure}[htb]
   \centering
   \centerline{\includegraphics[width=0.95\linewidth]{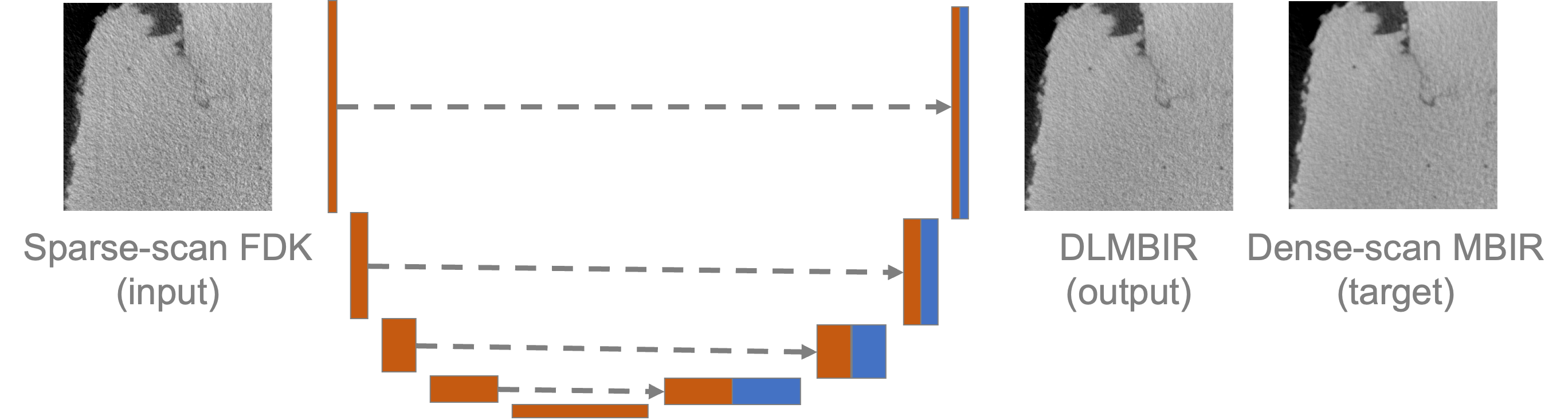}}
\vspace{-0.35cm}
 \caption{Illustration of the DLMBIR's U-Net architecture used for suppressing artifacts from an analytic reconstruction by learning a mapping between FDK and MBIR methods. 
 In U-Net, each block on the left side represents 2 \textit{convolution}+\textit{batch normalization}+\textit{ReLU} layers followed by a \textit{maxpooling} layer that halves the feature size. 
 The blocks on the right side represent \textit{transposed convolution} that doubles the feature size followed by 2 \textit{convolution}+\textit{batch normalization}+\textit{ReLU} layers. 
 The block height and width represent the size and number of features respectively.}
 \vspace{-0.2cm}
 \label{fig:Unet}
\end{figure}

\vspace{-0.25cm}
For the artifact and noise suppression network, we use a modified~\cite{rahman2021mbir} U-Net~\cite{ronneberger2015u} as shown in Fig.~\ref{fig:Unet}.
We 3D-printed components using a CAD design that consists of complex structures with a cylindrical base of 15 mm diameter, and 3 separate fixtures of different sizes, namely fins, inclines, and rods; and made with three materials -- steel, Aluminum Cerium (AlCe) and Inconel~\cite{snow2023observation}. 
We measured each part using a densely-sampled XCT scan, and then chose three of the data sets to generate training data for the neural networks.
%Three measurement was used to generate  training data for network, from three different materials.
Detailed measurement settings are provided in Table~\ref{Tab:Meas_setting}.

For generating training data, pairs of corrupted (with noise, BH and other artifact) and clean reconstruction data are obtained by using the FDK and MBIR algorithms respectively. 
To create training data, the samples were scanned densely using 500 
%(short-scan~\cite{ziabari2018model}) 
to 1000 
%(full-scan~\cite{ziabari2018model}) 
projections, and reconstructed using MBIR with BH correction applied (section~\ref{sec:BHCN}) to create various reference data.
Then, we sub-sampled the number of views by a factor of 3 or 4 for each data set, and reconstructed it using the FDK algorithm with BH correction applied to create noisy input data for training.
The U-Net-based network~\cite{rahman2021mbir}, is then trained on pairs of FDK and MBIR data discussed above to learn to suppress the noise and artifact from the FDK input and produce a high-quality output approximating MBIR method with a reduction in BH artifacts. 
We hypothesize that training the network on a variety of data with different settings, materials and sparsity conditions, allows the network to better adapt to in- and out-of-distribution data. 

\firsttable

\thirdtable

\vspace{-0.25cm}
\section{Results}
\label{sec:results}

We used the approach from~\cite{ziabari2023enabling, ziabari2022high} as the reference beam hardening correction method. 
This algorithm leverages the available CAD model of the part and physics-based model to extract BH parameters of an alloy used to correct for the beam hardening. 
But it is not universal and needs to be recalculated for every new alloy.
We used 3D-printed components of several alloys such as AlCe, steel, Inconel, and NiCo to test the proposed method. 
%The AM parameters and XCT scan parameters were similar to the ones mentioned in Sec. \ref{subsubsec:Training}.

The standard scan time and number of projections are typically determined by the expert operator using the XCT system to address the trade-off between measurement time (cost) and quality of the reconstruction. 
The typical duration used for XCT scanning of materials and geometry we discussed in this work, are between 30-60 minutes.%, or more for dense materials. 
The time is calculated by multiplying number of views, integration time and number of images per view as noted in Table~\ref{Tab:Meas_setting}.
With our method we are able to accelerate the XCT scans by up to 70\%  while obtaining reconstructions that are visually and quantitatively similar to the reference method. 
The description of the images that will be discussed in the next sub-sections can be found in Table~\ref{Tab:Image_description}.
\secondtable
\begin{figure}[htb]
  \centering
  \centerline{\includegraphics[width=\linewidth]{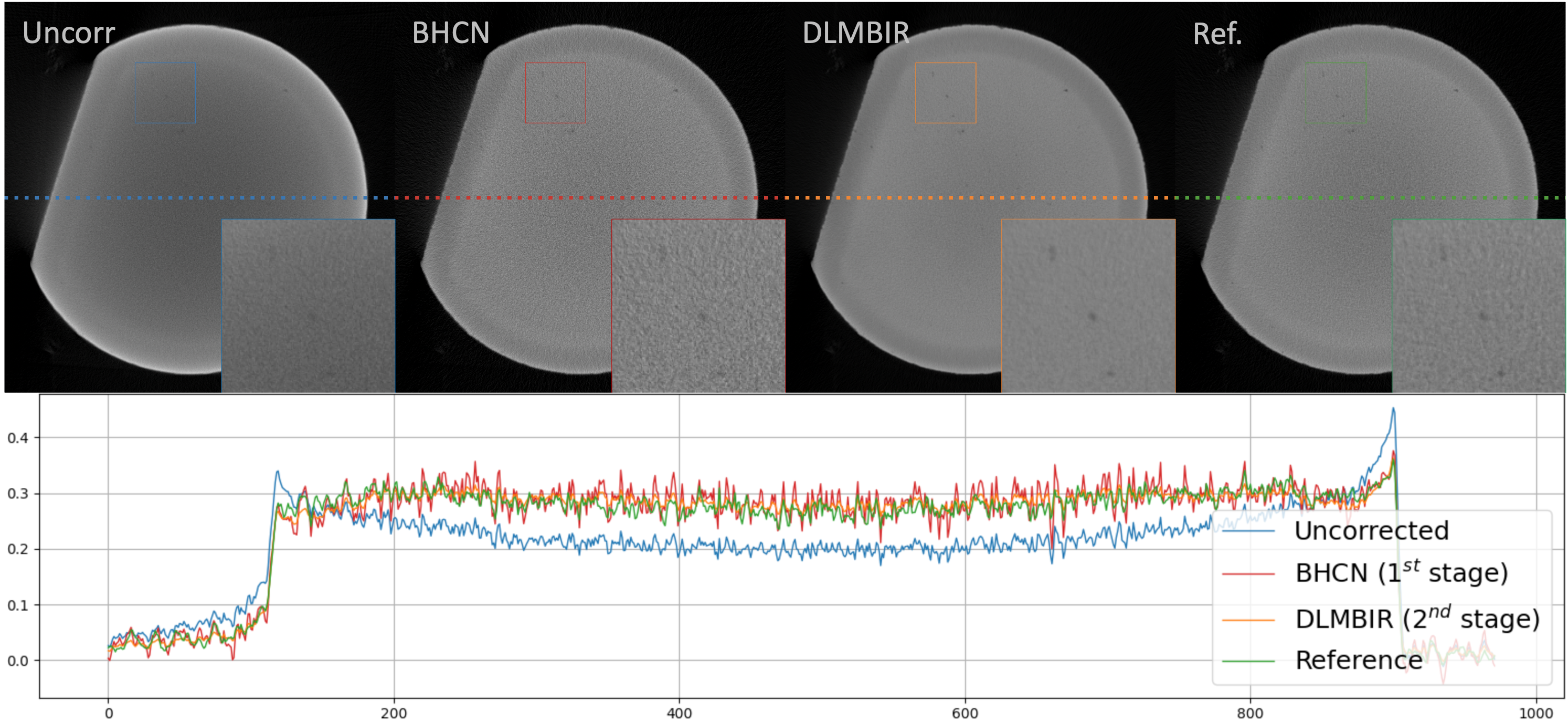}}
 \vspace{-0.35cm}
\caption{Different stages of artifact correction. (Left to right) Uncorrected FDK, BHCN-corrected ($1^{st}$ stage of our method), DL-MBIR-corrected ($2^{nd}$ stage of our method), and reference image; (Bottom) Profile plot. 
The inset image in the top row shows a small patch from the reconstructed volumes.
Notice that the proposed workflow results in reconstructions which have significantly less cupping artifacts and residual noise. 
}
\vspace{-0.5cm}
\label{fig:Correction_stages}
\end{figure}
\subsection{Qualitative analysis}
\begin{figure}[htb!]
  \centering
   \centerline{\includegraphics[width=\linewidth]{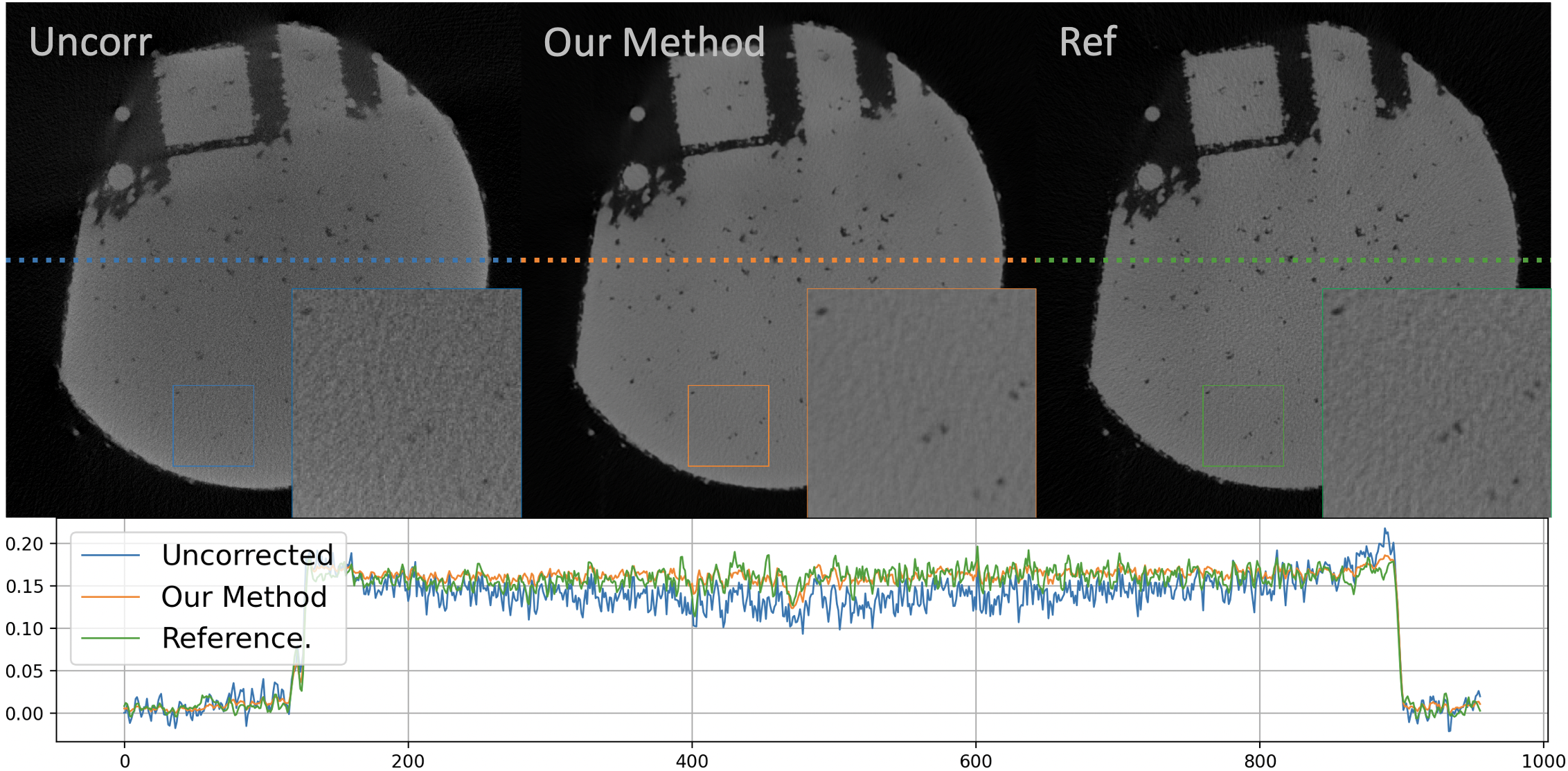}}
  \centerline{(a) AlCe}
    \centerline{\includegraphics[width=\linewidth]{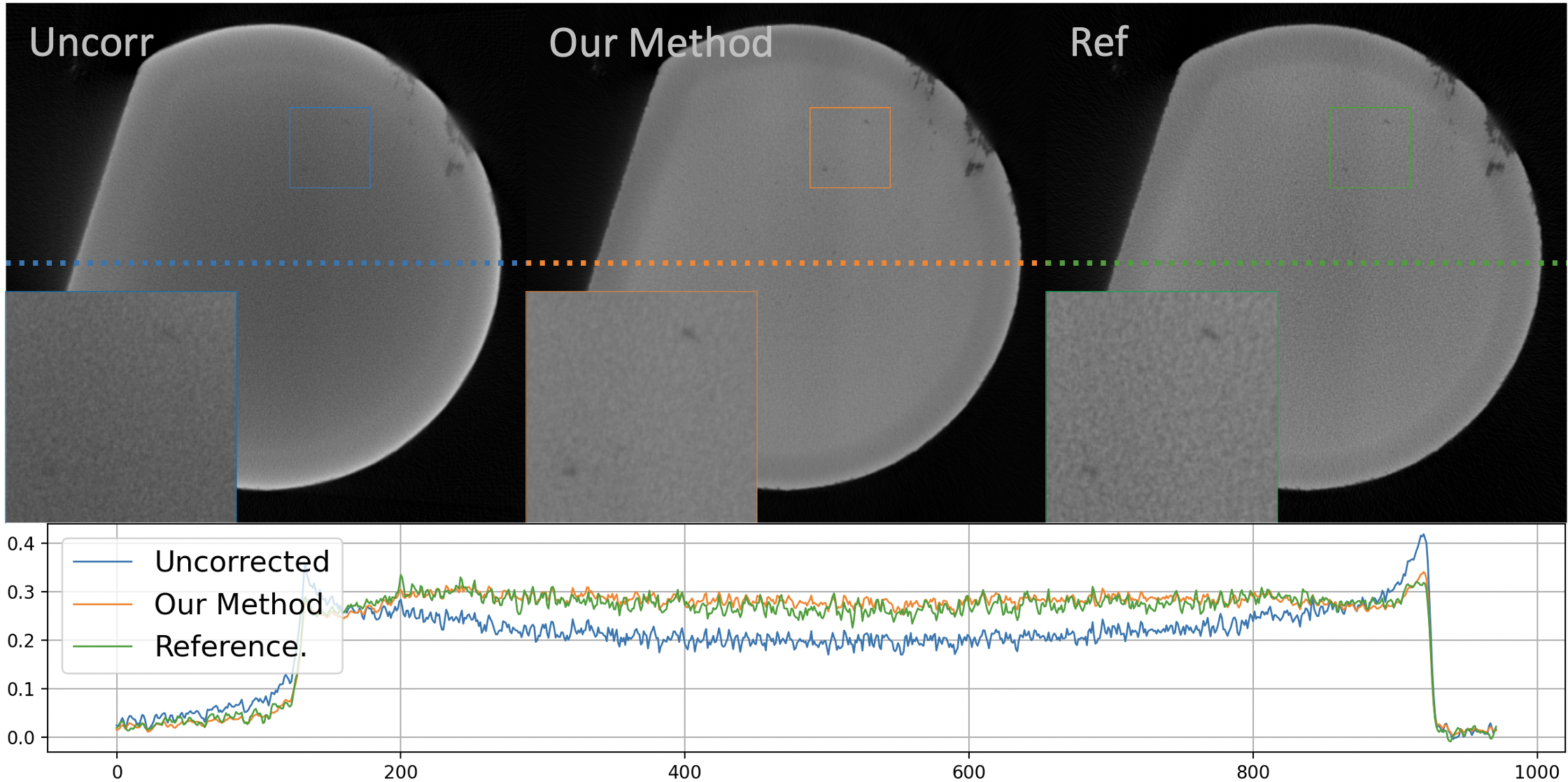}}
  \centerline{(b) Steel}
 \vspace{-0.35cm}
\caption{Testing on In-Dist data (See Table~\ref{Tab:Meas_setting}) from two different materials that were used during training time. (left to right) Uncorrected, our method, reference; Profile plot. 
The inset image of a smaller patch from the reconstruction demonstrates that defect contrast with the proposed method is higher than that of uncorrected, and comparable to MBIR. 
The profile plots clearly show reduction in BH with the proposed method.}
\vspace{-0.5cm}
\label{fig:In-distribution_BHCNN-DL-MBIR}
\end{figure}
\begin{figure}[h!]
  \centering
  \centerline{\includegraphics[width=\linewidth]{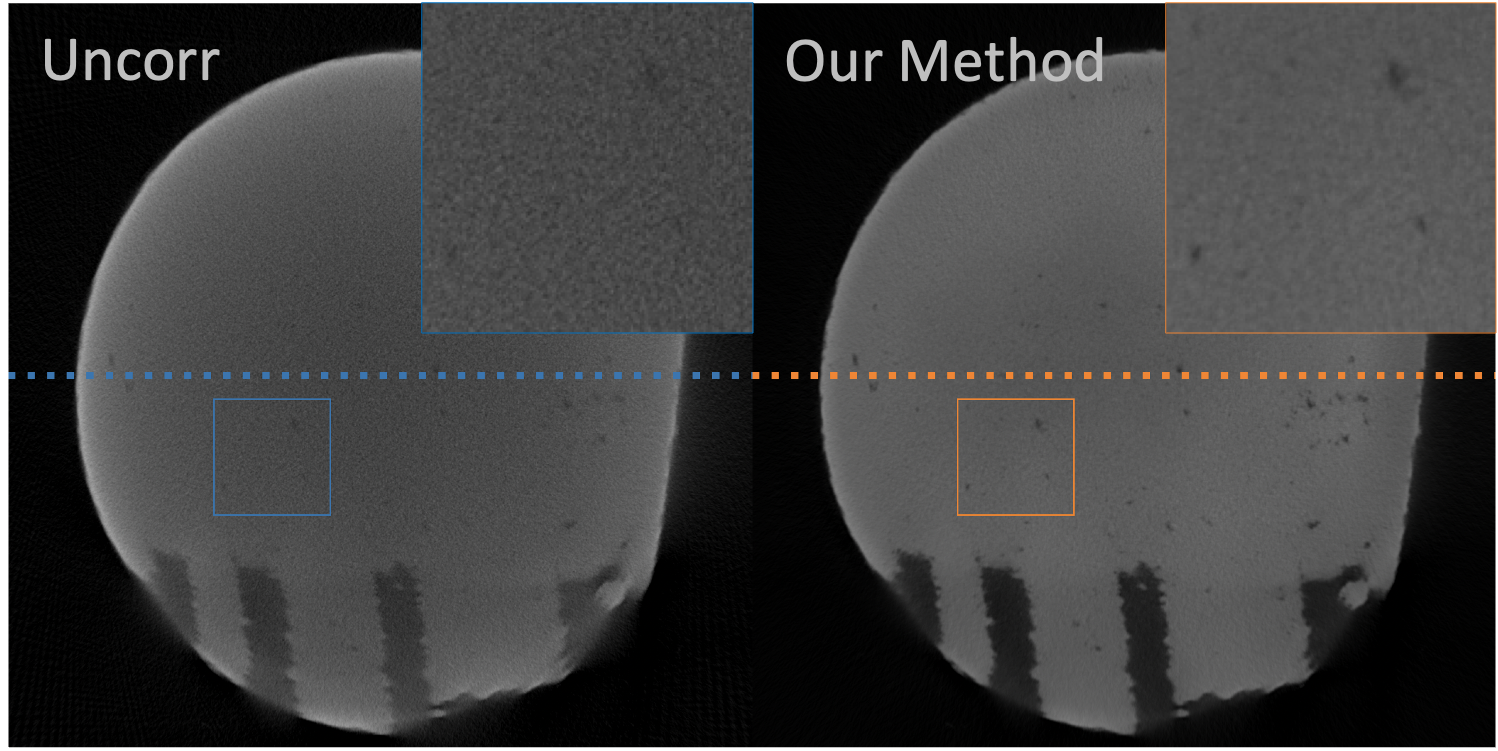}}
  \centerline{(a) Inconel}
  \centerline{\includegraphics[width=\linewidth]{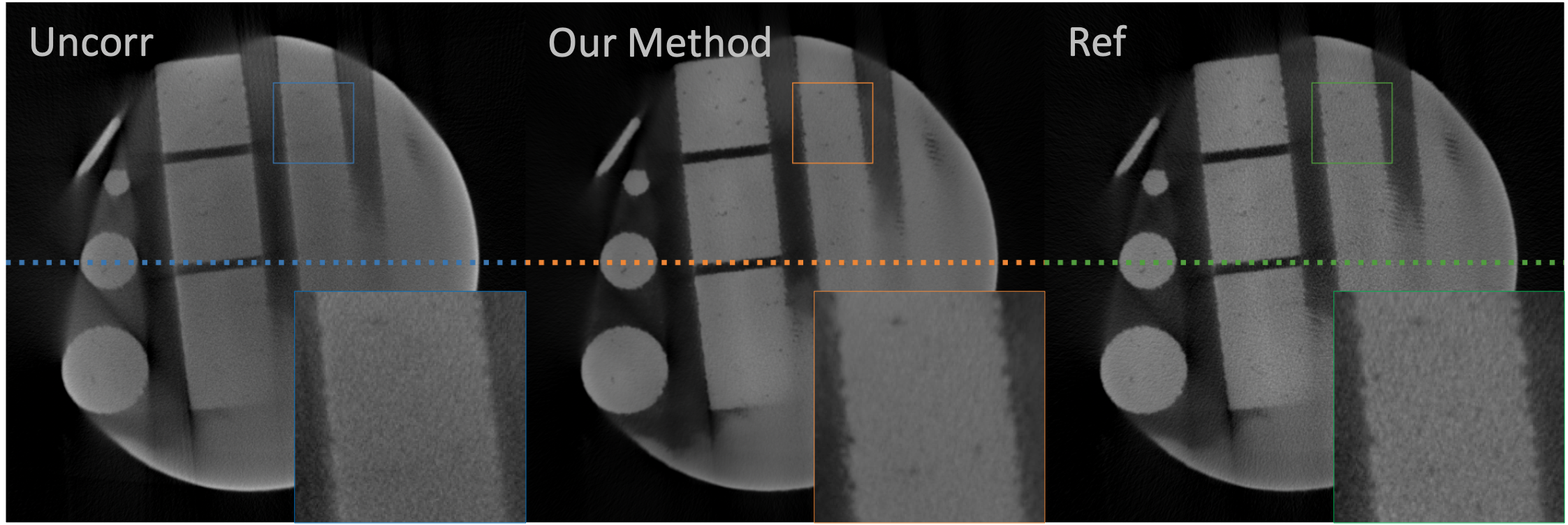}}
  \centerline{(b) NiCo}
\vspace{-0.35cm}
\caption{Testing on out-of-distribution data set. Please refer to Table~\ref{Tab:Meas_setting}. 
(Top subfigure left to right) Uncorrected, our method. 
(Bottom subfigure left to right) Uncorrected, our method, reference.
From the inset images of a small patch, we notice that the resulting reconstructions have fewer artifacts and can discern defects better than the baseline method. 
}
\vspace{-0.5cm}
\label{fig:Out-of-distribution_BHCNN-DL-MBIR}
\end{figure}
%To test the proposed method we printed components with alloys such as AlCe, NiCo, steel, Inconel, and AlSiMg. 
We trained the DLMBIR network for varying sparsity conditions and also for several alloys and expect the network not only to perform well on samples made of alloys that are in distribution (In-Dist.), but also to generalize on those that are  out of distribution (OOD). 
Table~\ref{Tab:Meas_setting} provides information on the training and testing alloys.
% We will also subject the network to scans of object with a severity of defects not encountered during training. 
In Fig.~\ref{fig:In-distribution_BHCNN-DL-MBIR}, it can be seen that in the left column, which is the uncorrected reconstruction, the defects have low contrast because of beam hardening and noise from sparse scan. 
The center column which is our method clearly shows an advantage and the defects are more noticeable with noise suppressed which should lead to more accurate detection of the defect structures.
The right column is the reference image.

We also used our method on data sets from alloy that were OOD with respect to the training data. 
Those include NiCo and Inconel, the latter of which was printed with a different 3D printer than the one used during training and scanned at a different setting. 
Fig. \ref{fig:Out-of-distribution_BHCNN-DL-MBIR}, shows the results for the OOD data highlighting the robustness of the method. 
While we haven't acquired the long dense scan required for creating a reference for test Inconel part, we can qualitatively observe the improved quality of reconstruction with respect to the input FDK reconstruction. 
Specifically, our method suppresses the BH artifact and noise, and highlights the defects. 
Our method also reduces artifacts on NiCo which is a significantly denser material compared to everything that was used during training.
However, in the NiCo case, we notice that our method misses some pores that are present in the reference image RoI, implying that we need a better training that makes DL-MBIR more general.
%
%\subsection{PSNR and SSIM}
Finally, we compare the deviation of the reconstructions w.r.t the reference using peak signal-to-noise ratio (PSNR) and structural similarity index measure (SSIM)~\cite{SSIM2004Wang} in Table~\ref{Tab:PSNR_SSIM}.
Our method has higher PSNR and SSIM than uncorrected across all the alloys which further demonstrates the effectiveness and robustness of our method.

\vspace{-0.25cm}
\section{Conclusion}
\vspace{-0.25cm}
In this paper we have presented a neural network-based workflow to enable high-quality XCT reconstructions of dense metal parts from industrial XCT systems while being able to reduce scan time and produce results in time comparable to traditional reconstruction but with higher quality. 
The proposed workflow allows for disentangling the beam hardening correction from the denoising (and sparse scan artifact reduction) process.
This in turn enables a robust material-agnostic BH correction, while suppressing the artifacts due to sparse-scanning strategy that is used to accelerate the scans and enable high-throughput inspections. 
Our experiments highlight that our workflow generalizes well to out-of-distribution data -- a key requirement to be able to deploy deep neural network-based solutions for real-world systems. 
Future work includes a more comprehensive training of the network, while experimenting across a wide range of materials and measurement settings.

\vspace{-0.25cm}
\section{Acknowledgement}
\vspace{-0.25cm}
\small
The 3D printing of Inconel data used for training was supported in a research collaboration with Dr. Brian Fisher from Raytheon Technologies.
Authors, would like to acknowledge Drs. Michael Kirka, Michael Sprayberry, Peter Wang, and Alex Plotkowski for providing components of different materials for testing the model.
The training and test parts were printed with different 3D printing systems at the Manufacturing Demonstration Facility (MDF) at Oak Ridge National Lab (ORNL).
%\newpage
%\balance
%\vfill\pagebreak
% -------------------------
\small
\bibliographystyle{IEEEbib}
\bibliography{refs}

\end{document}